\documentclass[11pt,a4paper,amsfonts]{article}

\usepackage{citesort} 

\usepackage{amsfonts}

\begin{document}

\title{Universally Coupled Massive Gravity, III:  dRGT-Maheshwari Pure Spin-$2$, Ogievetsky-Polubarinov and Arbitrary Mass Terms}

\author{J. Brian Pitts\footnote{Faculty of Philosophy, University of Cambridge; jbp25@cam.ac.uk.  Funded by the John Templeton Foundation, grant \#38761.}}
 	

\maketitle


\begin{abstract}

Einstein's equations were derived for a free massless spin-2 field using universal coupling in the 1950-70s by various authors; total stress-energy including gravity's served as a source for linear free field equations. A massive variant was likewise derived in the late 1960s by Freund, Maheshwari and Schonberg, and thought to be unique. How broad is universal coupling? In the last decade four 1-parameter families of massive spin-2 theories (contravariant, covariant, tetrad, and cotetrad of almost any density weights) have been derived using universal coupling. The (co)tetrad derivations included 2 of the 3 pure spin-2 theories due to de Rham, Gabadadze, and Tolley; those two theories first appeared in the 2-parameter Ogievetsky-Polubarinov family (1965), which developed the symmetric square root of the metric as a nonlinear group realization. One of the two theories was identified as pure spin-2 by Maheshwari in 1971-2, thus evading the Boulware-Deser-Tyutin-Fradkin ghost by the time it was announced. Unlike the previous 4 families, this paper permits \emph{nonlinear} field redefinitions to build the effective metric. By not insisting in advance on knowing the observable significance of the graviton potential to all orders, one finds that an \emph{arbitrary} graviton mass term can be derived using universal coupling. The arbitrariness of a universally coupled mass/self-interaction term contrasts sharply with the uniqueness of the Einstein kinetic term. One might have hoped to use universal coupling as a tie-breaking criterion for choosing among theories that are equally satisfactory on more crucial grounds (such as lacking ghosts and having a smooth massless limit). But the ubiquity of universal coupling implies that the criterion doesn't favor any particular theories among those with the Einstein kinetic term.

\end{abstract}

\vspace{.25in}
keywords: massive graviton; universal coupling; dRGT; spin 2; Vainshtein mechanism;  undiscovered public knowledge; counterfactual history


\section{Prelude:  Discoveries and Rediscoveries}

%
%
%

The project of deriving a relativistic gravitational theory using considerations such as an analogy to Maxwellian electromagnetism, the universal coupling of the gravitational field to a combined gravity-matter energy-momentum complex, and the requirement that the gravitational field equations alone (without the matter equations) entail energy-momentum conservation  was a part of Einstein's search for an adequate theory of gravity in 1913-15 \cite{EinsteinPapers8,EinsteinTrans4,Janssen,JanssenRenn}. 
This history was subsequently downplayed by Einstein, who thought that his strategy of principles had brought him success and needed to justify his persistent unified field theory quest 
 \cite{JanssenRenn,Feynman,vanDongenBook}. 
The physical ideas above later came to be associated with the (allegedly non-Einsteinian) field-theoretic approach to gravitation more commonly associated with particle physics, where it was brought to successful completion in the 1950s-60s \cite{Kraichnan,Gupta,Feynman,Weinberg65,Wyss,Deser}. Objections to such derivations \cite{PadmanabhanMyth} have been addressed \cite{MassiveGravity1,DeserRedux,Lasenby}.   

Until the 2000s it was believed that there was a unique massive generalization of General Relativity satisfying universal coupling \cite{FMS,DeserMass}. This theory has a ghost spin-$0$ with the same mass as the spin $2$, generating expectations of instability.  More recently two $1$-parameter families of massive theories were derived by the author \cite{MassiveGravity1}, being tied to covariant or contravariant tensors of various density weights. The ratio of spin-$0$ mass to spin-$2$ mass is determined by the index position and density weight.   In these theories the spin-$0$ mass is no larger than the spin-$2$ mass, so the ghost is always present.

Subsequent work  derived another $2$-parameter family of universally coupled massive gravities using a tetrad or cotetrad with arbitrary density weight \cite{MassiveGravity2}.  The spin-$0$ ghost mass ranged from $0$ to $\infty$ including both endpoints. %
 The four infinitely heavy spin-$0$ ghost cases were proposed  in tetrad form by Zumino \cite{ZuminoDeser,MassiveGravity2}. These theories are a small part of the 1965 Ogievetsky-Polubarinov $2$-parameter family of massive spin-$2$ gravities \cite{OP,OPmassive2}\footnote{The latter paper, though brief, presents the field equations without a typographical error present in the former.}  presumed at the time (before recognition of the Boulware-Deser ghost) to include pure spin $2$ theories simply by perturbatively having infinite spin-$0$ mass, as well as spin $2$-spin $0$ theories.  That paper exhibited a mild preference for the infinite spin-$0$ mass theories.

 Two of these theories were recently reinvented in a tensor-based formalism and were shown to avoid the Boulware-Deser nonlinear reappearance of a ghost avoided linearly \emph{via} the Fierz-Pauli mass term with infinite spin-$0$ mass  \cite{deRhamGabadadze,HassanRosenNonlinear}.   Thus in effect two of the de Rham-Gabadadze-Tolley-Hassan-Rosen ghost-free theories (out of the $3$ at fixed graviton mass, apart from mixtures)  were shown to satisfy universal coupling \cite{MassiveGravity2}, though the relation to nonlinear pure spin-$2$ was not pointed out. %

It turns out that the exact pure spin-$2$ character of one of these theories, the one using 
\begin{eqnarray} \sqrt{ g^{\mu\nu} \sqrt{-g}^2 \eta_{\nu\alpha} } \delta^{\alpha}_{\mu} =  \sqrt{-g} \sqrt{ g^{\mu\nu} \eta_{\nu\alpha} } \delta^{\alpha}_{\mu},
\end{eqnarray}
  was identified by Amar Maheshwari  in 1971-2 (submitted April 1971, published March 1972, according to the paper) \cite{MaheshwariIdentity}.  
  Maheshwari imposes a condition $np=-1$ relating the power $n$ of the densitized inverse metric $\tilde{g}^{\mu\nu}$---any real value (except perhaps $0$) using the binomial series expansion!---and the (negative) density weight $p$.  (Density weight is often defined  oppositely in Russian literature.)  
Proximately this condition presumably comes from equations 115 and 119 of Ogievetsky and Polubarinov (\cite{OP}); ultimately I find its motivation unclear, however.  Whatever its motivation, it didn't prevent Maheshwari from finding one of the pure spin-$2$ theories much discussed lately; in fact it helped, somewhat by chance perhaps.  Maheshwari knew his pure spin-$2$ theory  as  the Ogievetsky-Polubarinov theory $p=-2$ (density weight $2$ by western reckoning) $n=\frac{1}{2}$ (taking the square root of the resulting quantity). 
He shows it  to be exactly spin-$2$ by showing that the exact nonlinear field equations imply an algebraic condition of vanishing trace of the graviton potential.  This argument, one notes, makes use of neither the Stueckelberg mechanism, nor a tetrad, nor Hamiltonian methods with an ADM split, unlike much of the new literature.  The theory  fits the Fierz-Pauli form not merely at lowest order, but indeed to all orders. It thus gave over 40 years ago a  nonlinear completion of Fierz-Pauli gravity at least as far as avoiding the Boulware-Deser(-Tyutin-Fradkin) ghost is concerned\footnote{I thank Andrew Tolley for discussing how the theory fares with  the Vainshtein mechanism.}, something that was widely sought a few years ago.

The history of the Boulware-Deser(-Tyutin-Fradkin) ghost is also somewhat interesting.  It was diagnosed in papers received in May and June 1972 \cite{DeserMassLetter,DeserMass}.  In fact Tyutin and Fradkin made much of the Boulware-Deser ghost claim a bit earlier (the Russian version is dated March 1972 with submission in July 1971 \cite{TyutinMass}) in relation to the Ogievetsky-Polubarinov theories.  Perhaps one should speak of the Boulware-Deser-Tyutin-Fradkin ghost because the latter two authors only ascribe a nonlinear ghost to the Ogievetsky-Polubarinov theories---which is in fact true and was previously unrecognized for most of the linearly spin-$2$ cases in those theories, \emph{but not true for (at least) one of them} \cite{MaheshwariIdentity}.  
 The avoidance of the Boulware-Deser-Tyutin-Fradkin ghost was already achieved by Maheshwari by the time that its generic existence was supposedly shown.  It is difficult to show that anyone knew of Maheshwari's result besides Maheshwari, however:  the Springer web site for the paper shows exactly one citation from the 1980s, by Maheshwari himself, while Web of Science adds only a mention in a published bibliography in \emph{General Relativity and Gravitation} from 1978.

 This case of lost knowledge is an example of Hasok Chang's thesis that the history of science has neglected resources for the progress of contemporary science \cite{ChangWater}.  According to Deser, Izumi, Ong, and Waldron, the
\begin{quote}
recent revival is due to the (partial) resolution of an earlier, fatal, flaw:  mGR models
generically propagate a sixth, ghost-like, mode in contrast with the five physical
degrees of freedom (DoF) of their linearized, Fierz-Pauli (FP) counterparts. This
terminated interest in bimetric and fmGR models for four decades. \cite[footnotes suppressed]{DeserProblems}
\end{quote}
While there were exceptions, the general thrust of the passage is clearly correct, sociologically speaking, about the effects of the belief in the genericity of the Boulware-Deser-Tyutin-Fradkin ghost at least for more than three decades.  That ghost-free counterexamples exist is a deservedly celebrated result in the last 5 years.    But recalling that it was \emph{already shown by Maheshwari in 1971-2} that the ghost was not generic  even within the Ogievetsky-Polubarinov family, one is amazed in contemplating the difference between the actual history of massive gravity and a more rational counterfactual history (\emph{c.f.} \cite{NolanCounterfactualHistory}) that required only paying attention to published results.  One has a striking example of ``undiscovered public knowledge'' \cite{SwansonUndiscoveredPublicKnowledge}.
 With the original gesture toward nonperturbative effects also made by Vainshtein in 1972 \cite{Vainshtein}, there was no unresolved  significant barrier excluding hope for pure spin-$2$ theories---that is, nothing to block further research for decades.

 The recent historical narrative in literature on massive spin-$2$ gravity has given much prominence to the Boulware-Deser(-Tyutin-Fradkin) ghost---a theoretical problem allegedly showing that massive gravity is impossible---not just the van Dam-Veltman-Zakharov discontinuity (an empirical problem allegedly showing that massive gravity is not in fact correct).  Recognition of Maheshwari's work would have implied that the van Dam-Veltman-Zakharov discontinuity should have been seen as the real issue because the Boulware-Deser-Tyutin-Fradkin ghost problem was resolved before it was diagnosed.  Massive gravity wasn't known to be ``inconsistent,''  
but  only empirically falsified (or so it seemed).  Thus the new knowledge needed was how generally the Vainshtein mechanism worked, a result that gradually appeared during the 2000s.  Does  this case suggest a methodological reorientation  to  take literature searches more seriously, or even for departments to take the history of physics seriously in an institutionalized way  as a specialty  (as occurs at the University of Minnesota, and is not unknown in mathematics), rather than a hobby? 


\section{Introduction} 

The bulk of the paper is devoted to a novel and very general derivation of universal coupling and its applications to examples of interest, including the third (novel) dRGT-Hassan-Rosen theory and the entire Ogievetsky-Polubarinov family.  The basic idea is to avoid as long and as far as possible specifying the observable content of the gravitational potential $\gamma^{\mu\nu}$---that is, the precise details of its relation to the flat background metric $\eta_{\mu\nu}$ and the effective curved metric $g_{\mu\nu}$ seen by matter.  It turns out that the derivation works without making that specification of $\gamma^{\mu\nu},$ that is, giving the exact function of $\gamma^{\mu\nu}$ and $\eta_{\mu\nu}$ that determines the effective curved metric $g_{\mu\nu}.$ 
One can assume as an approximation a linear redefinition along the lines of $g_{\mu\nu}= \eta_{\mu\nu} + \sqrt{32\pi G} \gamma_{\mu\nu}$ and its densitized, contravariant, and contravariant densitized analogs \cite{Kraichnan,MassiveGravity2}, but now allowing for unspecified nonlinear terms as well.   By not insisting in advance on knowing the observable significance of the graviton potential exactly to all orders, one finds that an \emph{arbitrary} graviton mass term (including any algebraic self-interaction term)  can be derived using universal coupling (subject to mild invertibility assumptions).  One can always work backwards, and indeed do so easily, choosing an arbitrary nonlinear mass term and showing that it satisfies universal coupling for the right choice of graviton potential.  The physical meaning of the graviton potential is specified implicitly in building the effective curved metric $g_{\mu\nu}$ out of the initial flat metric $\eta_{\mu\nu}$ and graviton potential.  Working forwards from a specific nonlinear field redefinition to a universally coupled mass/self interaction term would  be much more difficult, but is not required. Working backwards by specifying a (universally coupled) mass/self interaction term, by contrast, is effortless.  
The phrase ``mass/self-interaction'' term is meant to imply the freedom of adding analogs of a $\phi^4$ interaction or the like in addition to the quadratic mass term.  The fact that one \emph{can} work backwards and have everything work is quite surprising, and depends crucially on the off-shell flexibility of the stress-energy tensor employed here, as embodied in the quantity $A_{\mu\nu}^{\rho\sigma} (\eta, \gamma)$ used below.

 Retaining flexibility of field definition till the end permits an enormous generalization of the theories that fit within the scope of universal coupling thus conceived.  
In particular, now the third dRGT-Hassan-Rosen theory also falls within the scope of universal coupling.  The other two, which already fit the universal coupling scheme in the tetrad formalism \cite{MassiveGravity2}, now fit within a metric-based formalism.  
Perhaps more importantly, an \emph{arbitrary mixture} of the three also is a universally coupled theory.  Likewise the entire $2$-parameter family of Ogievetsky-Polubarinov mass terms (including the first two dRGT-Hassan-Rosen theories) is now seen to satisfy universal coupling---or an arbitrary mixture of them, for that matter.  This paper, like their work \cite{OP}, uses a metric-based rather than tetrad-based formalism, in contrast to (\cite{MassiveGravity2}).  The one-parameter family of theories (at fixed spin-$2$ graviton mass) of Babak and Grishchuk are also included \cite{GrishchukMass,PetrovMass}, as is Visser's theory \cite{Visser}; these have ghosts.  The third dRGT-Hassan-Rosen theory, the Babak-Grishchuk theories and Visser's theory all  violate the pattern of $\sqrt{-g}$ + trace of a metric-like quantity + $0$th order term $\sqrt{-\eta}$ that held for earlier  examples \cite{OP,FMS,MassiveGravity1,PittsScalar,MassiveGravity2} and that one might have thought was obligatory.  
It is also striking that Ogievetsky and Polubarinov, by inventing nonlinear group realizations \emph{via} the binomial series expansion, in effect turned covariance \emph{vs.} contravariance from two isolated opposites into a continuous spectrum from contravariance ($n=1$) to covariance ($n=-1$), missing only the $0$ case.  It is possible  to speculate that a limiting procedure involving natural logarithms might fill that hole, $n=0.$  For the scalar case the analogous hole in the spectrum of density weights  has been filled \cite{PittsScalar}.  But such techniques are not needed here in any case.


It turns out, in fact, that \emph{any} algebraic mass/self-interaction term falls within the scope of universal coupling (subject perhaps to some mild invertibility requirements).  Whereas universal coupling leads uniquely to the Einstein tensor for the kinetic term (and/or higher-derivative terms built solely out of  $g_{\mu\nu}$), the algebraic terms are completely unrestricted, apart perhaps from some mild invertibility assumptions.  Consequently if one hoped that universal coupling could help to adjudicate  mass/self-interaction terms into categories of being more plausible or less plausible as a tie-breaker, that hope seems to be disappointed.  It is perhaps possible to regard some instances of the derivation as simpler and/or more natural than others---perhaps \emph{via} a principled restriction on terms proportional to the field equations in the definition of the stress-energy tensor; such considerations would seem to be the only way that universal coupling discriminates among mass/self-interaction terms.

One now realizes that good behavior as an effective theory ultimately counts for more than does the feature of universal coupling.  Most of this paper could have been written 40 years ago, before one thought in terms of effective field theory.  Hence in a sense one can use the results here to ascertain how people should have thought about massive gravity during the era when hardly anyone did so because of the supposed Boulware-Deser-Tyutin-Fradkin ghost (which Maheshwari had avoided in one case in advance).  If good effective field theoretic behavior conflicted with universal coupling, too bad for universal coupling.  Insofar as good effective field theoretic behavior leaves some choices unsettled, one might have hoped that universal coupling could be a further criterion for theory choice.  But that turns out not to be true either---unless some principled restriction narrowing down the freedoms of adding Euler-Lagrange derivatives to the stress-energy tensor and making field redefinitions can be devised.  Universal coupling is a nice idea, but turns out to be generic at least among theories with the Einstein kinetic term (an assumption no longer so taken for granted \cite{HinterbichlerKinetic}).


\section{Derivation of Einstein's Theory as Massless Spin-$2$}

The derivation of Einstein's equations as the field equations of a massless spin-$2$ theory is facilitated if one uses the metric stress-energy tensor \cite{Kraichnan,Anderson,Deser,SliBimGRG,MassiveGravity1}, usually attributed to Rosenfeld \cite{RosenfeldStress,GotayMarsden}.  Hilbert proposed defining stress-energy in terms of the variation of the action with respect to a metric; 
Rosenfeld adapted that derivation as a trick even in flat space-time by momentarily relaxing flatness in taking the variational derivative and then restoring flatness afterward.  
Such a trick  is related by identities to the canonical energy-momentum `tensor,' with (for fields transforming nontrivially under infinitesimal coordinate transformations, \emph{i.e.}, fields that are not scalars or pseudo-scalars) correction terms with identically vanishing divergence (`curls') and terms proportional to the equations of motion. This fact answers some of Padmanabhan's objections \cite{PadmanabhanMyth,MassiveGravity2}; one also notes the doubtfulness that the Hilbert action is the best regarding boundary terms, and the doubtfulness that the boundary terms matter if the formalism makes use of functional derivatives.

  The metric stress-energy tensor off-shell is rather sensitive to metric-dependent field redefinitions.  Thus the electromagnetic stress-energy tensor using Rosenfeld's recipe is different from the usual one from $A_{\mu}$ if one raises the index ($A^{\mu}$), densitizes  to weight $1$ ($\mathfrak{A}_{\mu} = \sqrt{-\eta}  A_{\mu}$) or some other weight, or both. How the action changes when one varies the metric $\eta_{\mu\nu}$  depends on whether $A_{\mu}$ or something else is held constant in the process.  
 Neither the canonical tensor nor the Belinfante-symmetrized tensor notices such field redefinitions.  Thus there is a combination of the metric stress-energy tensor and terms proportional to the field equations that is insensitive to metric-dependent field redefinitions.  However, this entity (identically equal to the Belinfante tensor) has second derivatives of the potentials, which is disappointing.  Hence two virtues, insensitivity to field redefinitions and the absence of second derivatives, compete for stress-energy tensors.  For electromagnetism one could argue that $A_{\mu}$ is the right choice for primitive field and hence $A^{\mu}$ is a wrong choice; thus invariance under field redefinitions might be discounted.  The apparatus of differential forms highlights those cases where covariant derivatives reduce to partial derivatives.  If one is taking a curl, a (possibly twisted) $1$-form (covector or axial covector) is much more convenient than anything else that one can build with the metric's help.    If one is taking a divergence, a (possibly axial) weight-$1$ tangent vector density is best.  One can take the dual of a tangent vector density of weight $1$ into a differential form, thus avoiding contravariant indices and densities.  But if one needs to take curls and divergences, or if one needs gradients, such a preferred choice will not exist.  Gravity is naturally described with a symmetric tensor (perhaps with density weight) \cite{WoodardSymmetricTetrad}, so derivative coupling to the metric is inevitable given the admission of tensors with respect to general coordinates.  Since the One True Field Definition seems not to exist for gravity, one might like a stress-energy tensor that is insensitive to field redefinitions.  But one might not want to have second derivatives.  With a forced choice between competing values, it seems prudent to be flexible about the notion of metric stress-energy tensor, permitting unspecified terms that vanish using the field equations.  Such flexibility will be displayed in the quantity $A^{\mu\nu}_{\alpha\beta}(\gamma,\eta)$ below.  One also thinks of topological field theories, for which the metric stress-energy tensor is identically $0$ off-shell \cite{Burgess}, as examples motivating not dogmatizing too much in favor of  the pure metric stress-energy tensor.
 
Many previous derivations of spin-$2$ gravities (massless or massive) found the need to treat the covariant and contravariant cases separately \cite{Kraichnan,MassiveGravity1,MassiveGravity2} (though the use of a near-continuum of covariance-contravariance by nonlinear group realizations has avoided that need previously \cite{OP}).   There are differences of detail---occasional minus signs, as well as an asymmetry due to the fact that an invariant action comes from a weight $1$ rather than weight $0$ Lagrangian density.  The payoff of such near-repetition was getting twice as many massive gravity theories.  However, this paper's admission of a nonlinear definition of the effective curved metric $g_{\mu\nu} $ (or inverse metric $g^{\mu\nu}$, or weight $1$ metric density $\mathfrak{g}^{\mu\nu}$, or weight $-1$ metric density $\mathfrak{g}_{\mu\nu}$, \emph{etc.})  obviates using both the contravariant and covariant derivations.  Either one will cover all cases.  I choose to use the contravariant case, partly to facilitate making closer contact with the classic  (and recently neglected) work of Ogievetsky and Polubarinov \cite{OP}, which expressed a $2$-parameter family of mass terms in terms of contravariant variables.


	For the massless theories, one assumes an initial infinitesimal invariance (up to a boundary term) of the free gravitational action.  For the later derivation of massive theories, the gauge freedom will be broken by a natural mass term algebraic in the fields, but the derivative terms will retain the gauge invariance.  This derivation follows (\cite{MassiveGravity1,SliBimGRG}) and so will be brief. 
 The gravitational potential is taken to be a contravariant symmetric tensor density field $\tilde{\gamma}^{\mu\nu}$ of density weight $l,$ for $l \neq \frac{1}{2}.$
 It will be convenient to use not the inverse flat metric itself, but its  weight $l$ densitized relative $\tilde{\eta}^{\mu\nu} = \eta^{\mu\nu}  \sqrt{-\eta} \: ^{l},$ where $l \neq \frac{1}{2}.$ 	For the massless theories, one assumes an initial invariance (up to a boundary term) of the free (linear in \emph{some} field) gravitational action $S_{f}$ under the infinitesimal gauge transformation 
$ \tilde{\gamma}^{\mu\nu} \rightarrow \tilde{\gamma}_{\mu\nu}  + \delta \tilde{\gamma}^{\mu\nu}$, where
\begin{eqnarray}
\delta \tilde{\gamma}^{\mu\nu} = \partial^{\mu} \tilde{\xi}^{\nu} + \partial^{\nu} \tilde{\xi}^{\mu} -l \eta^{\mu\nu} \partial_{\alpha} \tilde{\xi}^{\alpha} 
\end{eqnarray}
for $l \neq \frac{1}{2},$
$\tilde{\xi}^{\nu}$ being an arbitrary vector density field of weight $l.$ 
One can write equivalently
\begin{eqnarray}
\delta \tilde{\gamma}^{\mu\nu} = \tilde{\eta}^{\mu\alpha} \partial_{\alpha} \xi^{\nu} + \tilde{\eta}^{\nu\alpha} \partial_{\alpha} \xi^{\mu} -l \tilde{\eta}^{\mu\nu} \partial_{\alpha} \xi^{\alpha},  
\end{eqnarray}
which looks more like a Lie derivative of a flat metric (inverted and densitized) using its own covariant derivative. 
 	    For any $S_{f}$ invariant in this sense, a certain linear combination  the free field equations is identically divergenceless: 
\begin{equation}
\partial^{\mu} \left( \frac{\delta S_{f} } {  \delta \tilde{\gamma}^{\mu\nu} } - \frac{l}{2}\eta_{\mu\nu} \eta^{\sigma \alpha}  \frac{\delta S_{f} } {  \delta \tilde{\gamma}^{\sigma\alpha} }  \right) = 0.
\end{equation} 
This is the generalized Bianchi identity for the massless free field theory.

Local conservation of energy-momentum, which holds with the use of the Euler-Lagrange equations for gravity $\tilde{\gamma}^{\mu\nu}$ and matter $u$, can be written  as 
\begin{eqnarray}
\partial^{\nu} \left(
 \frac{\delta S}{\delta \tilde{\eta}^{\mu\nu} }  
-  \frac{l}{2}  \tilde{\eta}^{\alpha\beta} \tilde{\eta}_{\mu\nu} 
\frac{\delta S}{\delta \tilde{\eta}_{\alpha\beta} } \right)   = 0.
\end{eqnarray} 
It holds as a consequence of the field equations for gravity $\gamma$ and matter $u$.  Here $u$ represents an arbitrary collection of dynamical tensor (density) fields of arbitrary rank, index position,
and  weight; other geometric objects such as connections, projective connections, \emph{etc.} are also permissible.  It turns out for Einstein's theory that the matter field equations are not necessary to ensure conservation, but for the moment we do not assume Einstein's theory, and it will not hold for massive theories. For Einstein's equations these conservation laws remain true, but the conserved quantities have some peculiar  properties  due to the emergence of an additional gauge symmetry \cite{Band1,Quale,Grishchuk,BrazilLocalize}.  This is the bimetric version of the coordinate dependence of pseudotensors.  A reasonable interpretation, which works at least as well for pseudotensors in a single-metric formulation, is that there are more distinct conserved energy-momenta than expected, because there are more symmetries than expected \cite{EnergyGravity} and Noether's first theorem associates a conserved energy-momentum current with each translation symmetry.

Given the discussion above of the flexibility of metric energy-momentum tensors, it is useful to allow some gravitational field equation terms into the stress-energy tensor.  One could also allow matter field terms, but that seems unhelpful for present purposes.  The fact that realistic matter fields tend either to be of the Yang-Mills variety (hence not needing matter field equations to remove second derivatives  for the non-densitized  covector case) or to involve spinor fields (thus necessitating a refined discussion)  further suggests omitting matter field terms.

The universal coupling postulate is imposed in the form:
\begin{eqnarray}
\frac{\delta S}{\delta \tilde{\gamma}^{\mu\nu} } = \frac{\delta S_{f} }{\delta \tilde{\gamma}^{\mu\nu} } +   
\lambda \left(  \frac{\delta S}{\delta \tilde{\eta}^{\mu\nu} }    - A_{\mu\nu}^{\rho\sigma} (\eta, \gamma)  
\frac{\delta S}{\delta \tilde{\gamma}^{\rho\sigma} }   \right), 
\end{eqnarray}  
where $\lambda = -\sqrt{32  \pi G}$.  The sign of the coefficient of the stress-energy term is chosen in light of the contravariant index position.  Conceivably one could generalize the left side to permit some 
dependence on the graviton potential to multiply $ \frac{\delta S}{\delta \tilde{\gamma}^{\mu\nu} };$ no one worries whether $G^{\mu\nu}=0$ or $G_{\mu\nu}=0$ or $\mathfrak{G}^{\mu}_{\nu}=0$ are \emph{really} Einstein's field equations.  Even if that generality is a good idea, it merely adds a new term that can be absorbed into $A_{\mu\nu}^{\rho\sigma} (\eta, \gamma)$ and does not require any further accommodation.

Previously in arriving at GR or a $1$-parameter family of massive theories, a linear field redefinition was made to express the action in bimetric variables \cite{Kraichnan,Deser,MassiveGravity1}; instead of  $\tilde{\gamma}^{\mu\nu}$ and $\tilde{\eta}^{\mu\nu}$  one obtained  
$\tilde{g}^{\mu\nu}$ and $\tilde{\eta}^{\mu\nu}$ from  $\tilde{g}^{\mu\nu} = \tilde{\eta}^{\mu\nu}  + \lambda \tilde{\gamma}^{\mu\nu}.$ 
(Different choices of weight $l$ imply different handling of the trace of the gravitational potential.)
Tetrad definitions have also used linear field redefinitions \cite{DeserSupergravity,MassiveGravity2}. 
Much of innovation in this paper comes from relaxing that linear field redefinition by admitting unspecified nonlinear terms.  The old expression survives as the first-order approximation of the general expression that is given by a series:
\begin{eqnarray}
\tilde{g}^{\mu\nu} = \tilde{\eta}^{\mu\nu}  + \lambda \tilde{\gamma}^{\mu\nu} + O(\lambda \gamma)^2+ \ldots.
\end{eqnarray} 
Between the flexibility with the trace using the density weight and the new flexibility with unspecified nonlinear terms, this family of  bimetric changes of variables is a completely general series expansion.  Typically it will not be \emph{explicitly} invertible in closed form.  But it will be assumed tacitly that an inverse exists.  (If one thought that a mass-interaction term such as $sin^2$ was a good idea, such a person could perhaps argue that invertibility would be needed only for weak fields.)  Equating coefficients of the variations gives
\begin{eqnarray}
 \frac{\delta S}{\delta \tilde{\eta}^{\mu\nu}} |\tilde{\gamma} =   \frac{\delta S}{\delta
\tilde{\eta}^{\mu\nu}} |\tilde{g}  +  \frac{\delta S}{\delta \tilde{g}^{\rho\sigma}}   \frac{\partial \tilde{g}^{\rho\sigma} }{ \partial \tilde{\eta}^{\mu\nu} }
\end{eqnarray}
and 
\begin{eqnarray}   
 \frac{\delta S}{\delta \tilde{\gamma}^{\mu\nu}}  =    \frac{\delta S}{\delta \tilde{g}^{\rho\sigma}}  \frac{\partial \tilde{g}^{\rho\sigma} }{ \partial \tilde{\gamma}^{\mu\nu} }. 
\end{eqnarray}

Substituting this change of variables into the universal coupling postulate causes two terms to appear involving 
$  \frac{\delta S}{\delta \tilde{g}^{\rho\sigma}}$:
\begin{eqnarray}
\left[ \left(\delta^{(\mu}_{\alpha} \delta^{\nu)}_{\beta} + \lambda A_{\mu\nu}^{\alpha\beta} \right)  \frac{\partial \tilde{g}^{\rho\sigma} }{ \partial \tilde{\gamma}^{\mu\nu} }  - \lambda  \frac{\partial \tilde{g}^{\rho\sigma} }{ \partial \tilde{\eta}^{\mu\nu} } \right]  \frac{\delta S}{\delta \tilde{g}^{\rho\sigma}}
=\frac{\delta S_{f}}{\delta \tilde{\gamma}^{\mu\nu}} +  \lambda \frac{\delta S}{\delta \tilde{\eta}^{\mu\nu}} | \tilde{g}.
 \end{eqnarray}
For the old linear bimetric field redefinitions, the terms involving  $\frac{\delta S}{\delta \tilde{g}^{\rho\sigma}}$ cancelled out (with $A_{\mu\nu}^{\alpha\beta}=0$).  With the new nonlinear field redefinition, those terms still cancel out to $0$th order in $\lambda \gamma$ in their coefficients (as long as $A_{\mu\nu}^{\alpha\beta}$ also vanishes to $0$th order).  The higher order terms due to the nonlinearity of the field redefinition typically do not cancel, because $ \frac{\partial \tilde{g}^{\rho\sigma} }{ \partial \tilde{\eta}^{\mu\nu} } \neq \frac{\partial \tilde{g}^{\rho\sigma} }{ \partial \tilde{\gamma}^{\mu\nu} }$. But such terms can be cancelled by a judicious choice of $A_{\mu\nu}^{\alpha\beta}.$  Let us choose it accordingly and achieve
\begin{eqnarray}  \left(\delta^{(\mu}_{\alpha} \delta^{\nu)}_{\beta} + \lambda A_{\mu\nu}^{\alpha\beta} \right)  \frac{\partial \tilde{g}^{\rho\sigma} }{ \partial \tilde{\gamma}^{\mu\nu} }  - \lambda  \frac{\partial \tilde{g}^{\rho\sigma} }{ \partial \tilde{\eta}^{\mu\nu} }  =0.
\end{eqnarray}
Universal coupling then takes the form 
\begin{eqnarray} 
\frac{\delta S_{f}}{\delta \tilde{\gamma}^{\mu\nu}} =  -  \lambda \frac{\delta S}{\delta \tilde{\eta}^{\mu\nu}} | \tilde{g},
\end{eqnarray}
just as in the old derivation with a linear field redefinition.

The  free field generalized Bianchi identity and the universal coupling postulate together imply that 
\begin{eqnarray}
\partial^{\nu} \left(
 \frac{\delta S}{\delta \tilde{\eta}^{\mu\nu} }  | \tilde{g}
-  \frac{l}{2}  \tilde{\eta}^{\alpha\beta} \tilde{\eta}_{\mu\nu} 
\frac{\delta S}{\delta \tilde{\eta}_{\alpha\beta} } | \tilde{g}  \right)   = 0
\end{eqnarray} 
as an identity (off-shell). That is equivalent to \begin{eqnarray}
\partial_{\nu}  \frac{\delta S}{\delta {\eta}_{\mu\nu} }  | \tilde{g}  = 0
\end{eqnarray} 
identically.  
  In words, whatever dependence the full interacting action has on the flat background metric after the effective curved metric has been introduced, it contributes to the metric stress-energy tensor 
only a symmetric entity with identically vanishing divergence.  
     The quantity $\frac{\delta S}{ \delta \eta_{\mu\nu}} |\tilde{g}$ thus  has the form 
\begin{eqnarray}   
\frac{\delta S}{ \delta \eta _{\mu\nu}} |\tilde{g} = \frac{1}{2}  \partial_{\rho} \partial_{\sigma} \left( 
{\mathcal{M}}
^{[\mu\rho][\sigma\nu]} +   {\mathcal{M}}
^{[\nu\rho][\sigma\mu]} \right)  + B \sqrt{-\eta} \eta^{\mu\nu}
\end{eqnarray}
\cite[pp. 89, 429]{Wald} \cite{Kraichnan,Kraichnan2}, where ${\mathcal{M}} ^{\mu\rho\sigma\nu}$ is a tensor
density of weight $1$ and $B$ is a constant.  This result follows from the converse of
Poincar\'{e}'s lemma in Minkowski spacetime.
Thus the action consists of an arbitrary part that does not depend on the flat metric at all, $$S_{1}[\tilde{g}^{\mu\nu},u],$$
and a piece $S_2$ made of constants, flat metric Riemann tensor terms (to give curls in the stress-energy tensor),  and boundary terms:
\begin{eqnarray}   
S_{2} = \frac{1}{2} \int d^{4}x R_{\mu\nu\rho\sigma} (\eta)
{\mathcal{M}} ^{\mu\nu\rho\sigma} 
 + \int  d^{4}x \alpha^{\mu},_{\mu} + 2 B \int d^{4}x \sqrt{-\eta}. 
\end{eqnarray}
 $S_{2}$ contains all the ineliminable dependence on the background metric, and makes no contribution to the
 field equations.  Taking the simplest choice of $S_{1}$ gives the Hilbert action for Einstein's equations, along with a cosmological constant, though one could admit higher derivatives from powers of $g_{\mu\nu}$'s Riemann tensor.  The choice from the allowed values of $l$ makes no difference in the massless case. Neither does the nonlinearity of the field redefinition.  Thus one arrives at Einstein's equations uniquely within the realm of second-order partial differential equations.  If one wants to admit higher powers and/or covariant derivatives of the Riemann tensor for the curved metric, they are available.  That feature is due to the feature, inspired by Kraichnan \cite{Kraichnan} (in contrast to  some other derivations \cite{Feynman,OP,Deser}) of maintaining great generality of field definitions and avoiding gratuitous explicitness.  Kraichnan's derivation was admired by Bryce DeWitt (\cite{DeWittDissertation} \cite[p. xiv]{Feynman} and personal communication 15+ years ago).

If one thinks that the use of this term $ \frac{1}{2} \int d^{4}x R_{\mu\nu\rho\sigma} (\eta) {\mathcal{M}} ^{\mu\nu\rho\sigma} (\eta_{\mu\nu}, g_{\mu\nu}, u ) $ is too clever to be invented without knowing Einstein's theory in advance and thus cheating \cite{PadmanabhanMyth}, one could simply add to the metric stress-energy tensor a curl term $ \frac{1}{2}  \partial_{\rho} \partial_{\sigma} \left( {\mathcal{M}}
^{[\mu\rho][\sigma\nu]} +   {\mathcal{M}} ^{[\nu\rho][\sigma\mu]} \right)  + B \sqrt{-\eta} \eta^{\mu\nu} $ by hand, as has been noted \cite{MassiveGravity1}. The 4-divergence $\alpha^{\mu},_{\mu}$ resolves worries \cite{PadmanabhanMyth} about getting terms not analytic in the coupling constant $\lambda.$ (It is unclear that the Hilbert action is best in any event, given its badly behaved conservation laws \cite{PetrovKatz2}.)   If one finds it disturbing to momentarily relax flatness of $\eta_{\mu\nu}$ to take derivatives and then restore flatness, one can use the Rosen-Sorkin Lagrange multiplier trick, which makes the flatness of the flat metric a consequence of the variational principle \cite{RosenMultiplier,SorkinScalar}.

One might wonder whether this style of derivation of General Relativity from conservation laws bears any close resemblance to Noether's theorems.  In fact the derivation is in effect an adaptation to the Rosenfeld (variation of flat metric) stress-energy tensor a converse mentioned already by Emmy Noether and related to Felix Klein's work \cite{PetrovNoether}.  As Brading and Brown explain, in Einstein's theory  ``\ldots the Noether current can be expressed as consisting of a term which vanishes on-shell\ldots and a term whose divergence vanishes identically.'' \cite{BradingBrownSymmetries}
Such conservation laws, justly or not, have been called ``improper.'' 
 But Noether herself included a converse:
\begin{quote} 
 If [action] $I$ admits of the displacement group, then the energy relationships
become improper if and {\bf only if} $I$ is invariant with respect to an infinite group containing
the displacement group as subgroup. [footnote suppressed]  \cite[emphasis added]{Noether}  \end{quote} 
Given this link to Noether's theorem and the Belinfante-Rosenfeld relation between canonical and metrical stress-energy, one could envisage a parallel derivation of Einstein's theory without the flat metric tensor, albeit much less convenient.  It is not coincidental that universal coupling derivations for massive scalar gravity using the canonical tensor have led only to a single theory \cite{FreundNambu}, because one needs to use the freedom to add a curl to the canonical tensor to accommodate second derivatives except in one case \cite{PittsScalar}.



\section{Derivation for Contravariant Tensor Density Potential: Massive Case}

In the classic paper that did so much to end work on massive gravity for decades, Boulware and Deser showed appreciation for the principle of universal coupling in choosing among mass terms (after more crucial distinctions such as ghost \emph{vs.} no ghost had been settled): 
\begin{quote} 
It is of interest, both for its own sake and to illustrate the physical requirements we impose on finite-range models, to consider one version of a ``ghost'' theory in detail. The most appealing model is the one which carries over from general relativity the property that the source of the linearized field is the full stress tensor of the field itself.$^{24}$
\cite{DeserMass} \end{quote} 
(The footnote refers to the Freund-Maheshwari-Schonberg paper \cite{FMS} and discusses that theory's ghost.) 
Boulware and Deser, like Freund, Maheshwari and Schonberg, thought that universal coupling gave a unique answer and that the principle was appealing, appealing enough that the theory satisfying it would be better than theories not satisfying it, other things being equal.  Boulware and Deser thought that the resulting ghost theory was bad, unlike Freund, Maheshwari and Schonberg at the time. It has been shown in recent years that the result isn't unique \cite{MassiveGravity1} and that it doesn't lead purely to ghost theories at linear order \cite{MassiveGravity2} (or, as it turns out, even nonlinearly).  Here I show that universal coupling doesn't rule out anything if one makes enough use of field redefinitions.   Consequently all ghost-free theories that were not included before are included now (as well as all ghost or tachyon theories).

For the massive generalization of this contravariant derivation, the choice of density weight $l$ makes a difference. So will the nonlinearities.  While it is easy enough to run the derivation forward for linear field redefinitions \cite{MassiveGravity1}, that looks typically difficult at best with a nonlinear field redefinition.  Fortunately, one can always work backwards, showing that basically any graviton mass/(algebraic) self-interaction term is universally coupled for some (typically nonlinear) field redefinition!  If one does not insist on knowing in advance exactly to all orders how the graviton potential relates to experience, then one is rewarded with a super-abundant harvest of new universally coupled massive gravities.  A field (re)definition is a conventional choice: there is no fact of the matter about whether gravity is \emph{really}  $$\frac{g_{\mu\nu} - \eta_{\mu\nu} }{\sqrt{32 \pi G}}$$ or $$\frac{ \eta^{\mu\nu}\sqrt{-\eta} -\mathfrak{g}^{\mu\nu} }{\sqrt{32 \pi G}}$$ or the like, though some choices are more convenient or simpler than others, or are so for particular purposes (not necessarily the same choice for every purpose---witness the utility of $\mathfrak{g}^{\mu\nu}$ for wave equations \cite{Papapetrou} and $g_{\mu\nu}$ for a Hamiltonian treatment with simple primary constraints).  Thus nothing so important as whether a theory satisfies universal coupling should depend on there being a True choice.  One can admit them all. In building the massive theories using kinetic terms from the massless theories, I do not explore the interesting question of generalized kinetic terms \cite{HinterbichlerKinetic}.

To find the massive theories, one assumes a free theory with action $S_{f}$ (quadratic in the gradient of some field variables) consisting of a kinetic part  $S_{f0}$, which plays exactly the role that $S_f$ played in the massless case, and a mass term $S_{fm}$ quadratic in those same unknown field variables.   One seeks the full self-interaction gravitational  theory's action $S$ by assuming that it also splits into two parts $S_{0}$ and $S_{ms}.$   $S_{f0}$ leads to $S_0$ exactly as $S_f$ led to $S$ above, while the new free graviton mass term $S_{f0}$ leads to $S_{ms}$.  A key difference is that the erstwhile free field Bianchi identity applies to the massless piece, whereas the mass terms have no such restriction. This difference explains why the kinetic term is unique (assuming at most second derivatives in the field equations), whereas the mass term is arbitrary.

	Requiring $S_{f0}$ to change only by a boundary term under the infinitesimal variation  
\begin{eqnarray}
\delta \tilde{\gamma}^{\mu\nu} = \partial^{\mu} \tilde{\xi}^{\nu} + \partial^{\nu} \tilde{\xi}^{\mu} -l \eta^{\mu\nu} \partial_{\alpha} \tilde{\xi}^{\alpha} 
\end{eqnarray}
for $l \neq \frac{1}{2}$
implies the identity
\begin{eqnarray}
\partial^{\mu} \left( \frac{\delta S_{f0} } {  \delta \tilde{\gamma}^{\mu\nu} }  - \frac{l}{2}\eta_{\mu\nu} \eta^{\sigma \alpha}  \frac{\delta S_{f0} } {  \delta \tilde{\gamma}^{\sigma\alpha} }  \right) = 0.
\end{eqnarray}
Again we postulate universal coupling, make an unspecified nonlinear field redefinition, and arrive at 
\begin{eqnarray} 
\frac{\delta S_{f}}{\delta \tilde{\gamma}^{\mu\nu}} =  -  \lambda \frac{\delta S}{\delta \tilde{\eta}^{\mu\nu}} | \tilde{g}.
\end{eqnarray}

Making use of the split of both the free and interacting actions into massless and mass parts, one gets two universal coupling conditions.  The massless part satisfies Bianchi identities and so depends on the flat metric $\eta_{\mu\nu}$ only \emph{via} terms that do not affect the field equations; the effective curved metric density has swallowed much of the original dependence on the flat geometry.  Thus one gets an action for Einstein's equations (perhaps with cosmological constant or higher derivatives) for the massless piece.  
The  free and interacting mass terms are related by the condition  
\begin{eqnarray}
  \frac{\delta S_{fm}}{\delta \tilde{\gamma}^{\mu\nu}}   = - \lambda \frac{\delta S_{ms} }{\delta \tilde{\eta}^{\mu\nu}} | \tilde{g}.
 \end{eqnarray}
$S_{fm}$ is quadratic in the gravitational potential: 
$$ \frac{\delta S_{fm}}{\delta \tilde{\gamma}^{\mu\nu}}  = -m^2 \sqrt{-\eta}  \tilde{\gamma}^{\alpha\beta} ( \tilde{\eta}_{\alpha \mu} \tilde{\eta}_{\beta \nu}  + b \tilde{\eta}_{\alpha \beta} \tilde{\eta}_{\mu \nu}).$$ 
$m$ is the mass of the spin-$2$ graviton.  For those theories that have a spin-$0$ graviton linearly as well, $b$ accommodates the spin-$0$ mass, which need not be the same as $m.$  The case value of $b$ that gives the Pauli-Fierz pure spin-$2$ free mass term depends on the density weight $l$.

When one had a linear field redefinition into bimetric variables \cite{MassiveGravity1}, it was straightforward to work forward, changing $ \frac{\delta S_{fm}}{\delta \tilde{\gamma}^{\mu\nu}}$ into bimetric variables, and arrive at affine dependence on $\tilde{g}^{\mu\nu}$ in the mass term apart from an algebraic weight $1$ expression built purely in the metric (a formal cosmological constant term), as in the $n=\pm 1$ Ogievetsky-Polubarinov theories  and the Freund-Maheshwari-Schonberg theory (the first massive gravity theory derived by universal coupling):
$$S_{ms} =  \frac{m^2}{16 \pi G} \int d^{4}x ( \sqrt{-g}[2l-1] - \sqrt{-\eta}[2l-3] - \frac{1}{2}\sqrt{- \eta} \, \tilde{g}^{\mu\nu} \tilde{\eta}_{\mu\nu}) $$   
for $l \neq \frac{1}{2}.$    A similar expression for a different family of theories was derived using covariant fields with density weight $-l$:
$$S_{ms} =  \frac{m^2}{16 \pi G} \int d^{4}x ( \sqrt{-g}[1-2l] + \sqrt{-\eta}[2l+1] - \frac{1}{2}\sqrt{- \eta} \, \tilde{g}_{\mu\nu} \tilde{\eta}^{\mu\nu}) $$ 
for $l \neq \frac{1}{2}.$  
Similar results appeared with a (co)tetrad density derivation \cite{MassiveGravity2}.  For the covariant tetrad density cases with weight $-w$
one found
$$ S_{ms} =  \frac{m^2}{8 \pi G} \int d^{4}x ( \sqrt{-g}[1-4w] + \sqrt{-\eta}[4w+3] - \sqrt{- \eta} \, \tilde{g}_{\mu}^A \tilde{\eta}^{\mu}_A)
$$ for $w \neq \frac{1}{4}.$
For the contravariant tetrad density cases of weight $w$ one found
$$ S_{ms} =  \frac{m^2}{8 \pi G} \int d^{4}x ( \sqrt{-g}[4w-1] + \sqrt{-\eta}[5-4w] - \sqrt{- \eta} \, \tilde{g}^{\mu}_A \tilde{\eta}_{\mu}^A) 
$$
for $w \neq \frac{1}{4}.$ A weight $-w$ cotetrad implies a weight $-l=-2w$ metric; a weight $w$ tetrad implies a weight $l=2w$ contravariant metric.    Avoiding tachyons imposes restrictions in each case \cite{OP,MassiveGravity1,MassiveGravity2}.  The (co)tetrad theories permit the spin-$0$ graviton to be heavier than the spin-$2$, even infinitely heavy, thus getting rid of the ghost at least to linear order; these are Zumino's theories \cite{ZuminoDeser}.  Of those 4 theories, $2$ of them are exactly ghost-free, while the other two suffer from the Boulware-Deser-Tyutin-Fradkin ghost nonlinearly. Zumino seems to have assumed that getting rid of the ghost to lowest order kept it away permanently.

Because it seems impossible to think of other linear geometric objects to represent space-time geometry and gravity, one might have conjectured that only these four families of theories were universally coupled.  But the unspecified nonlinear field redefinition simultaneously effects two changes:  it renders working forward typically difficult or impossible instead of easy, and it makes working backward always easy instead of typically difficult or impossible.  To work forward is to specify completely a nonlinear bimetric field redefinition and infer the interacting mass term $S_{ms}$ and hence the whole theory.  To work backward is to specify $S_{ms}$ and infer the bimetric field redefinition.  The mass term  is not in danger of already needing to know the answer in advance.  Because every answer $S_{ms}$ is permissible, working backwards is not cheating.

Returning to the new derivation, one has 
\begin{eqnarray}
 - m^2 \sqrt{-\eta}  \tilde{\gamma}^{\alpha\beta} ( \tilde{\eta}_{\alpha \mu} \tilde{\eta}_{\beta \nu}  + b \tilde{\eta}_{\alpha \beta} \tilde{\eta}_{\mu \nu}) =  \frac{\delta S_{fm}}{\delta \tilde{\gamma}^{\mu\nu}}   = - \lambda \frac{\delta S_{ms} }{\delta \tilde{\eta}^{\mu\nu}} | \tilde{g}.
 \end{eqnarray}
One can now read the equation backwards:  instead of assuming that one knows precisely the empirical meaning of the left side (using some specific nonlinear field redefinition) and inferring the universally coupled mass term on the right side, one can read it as defining  $\tilde{\gamma}^{\mu\nu}$ in terms of $\tilde{g}^{\mu\nu}$ and $\tilde{\eta}^{\mu\nu}.$  This expression gives the inverse of the nonlinear bimetric field redefinition.  
Because so far the nonlinear field redefinition has been specified only to linear order, this equation cannot fail to hold for any reasonable mass term $S_{ms}$; indeed $S_{ms}$ ought to be just $S_{fm}$ to that order.    
One is asking the question, ``what should one mean by the gravitational potential in order to arrive at mass/self-interaction term $S_{ms}$ \emph{via} universal coupling?''   One would have expected that typically no answer to this question would exist.  But in fact generically there is an answer.


\section{Bianchi Identity for Full Theory and Vanishing Divergence Conditions}

If one wants to eliminate spin-$1$ and one spin-$0$ excitations, one desires a vanishing coordinate divergence (or covariant divergence with respect to $\eta_{\mu\nu}$) \cite{OP,MaheshwariIdentity}.  If one takes the $g$-covariant divergence of the gravitational field equations $\frac{ \delta S}{\delta \tilde{g}^{\mu\nu}}=0$ or the like, then one seems to be stuck with curved $g$-Christoffel symbols.  Even in the best case for taking a divergence, using $g_{\mu\nu}$ as the field variable and getting Euler-Lagrange equations with density weight $1$ and contravariant character $ \frac{ \delta S}{\delta g_{\mu\nu}}=0$, one still has one Christoffel symbol term remaining.  It might be tempting to think that such a covariant divergence cannot be written as a coordinate or flat covariant divergence, but in fact it can.  Neglecting matter in this section for simplicity of exposition, one has a gravitational action that is a scalar.  Expanding out the Lie derivatives of the metrics and remembering that it is a contravariant vector that describes a small coordinate transformation, one gets the generalized Bianchi identity 
\begin{eqnarray} 
g_{\alpha\nu}  \nabla_{\mu} \frac{ \delta S }{ \delta g_{\mu\nu} }   +  \eta_{\alpha\nu}  \partial_{\mu} \frac{ \delta S  }{ \delta \eta_{\mu\nu} } |g=0.
\end{eqnarray}  
In this way one relates a $g$-covariant divergence of the derivative of the mass term with respect to $g_{\mu\nu}$, to the $\eta$-covariant divergence of the derivative of the mass term with respect to $\eta_{\mu\nu}$.  Off-shell one can use this identity either for the whole action or for just the mass term.  For the whole action, the gravitational field equations $\frac{ \delta S }{ \delta g_{\mu\nu} }=0$ annihilate the first term, so the second term gives an on-shell vanishing divergence, which provides the desired conditions eliminating spin-$1$ and a spin-$0$.  If matter were present, one would need to use its Euler-Lagrange equations as well.  In any case, with everything on-shell one gets 
\begin{eqnarray}   
 \eta_{\alpha\nu}  \partial_{\mu}   \frac{ \partial \mathfrak{L}_{ms}  }{\partial \eta_{\mu\nu} } |g=0;
\end{eqnarray}  
the $|g$ is now added as a reminder of the use of bimetric variables, so that the curved metric rather than the graviton potential is the other independent variable.  
For theories in the form $\sqrt{-g}$ piece + $\sqrt{-\eta}$ piece + trace of metric-like entity, only the term involving trace of the metric-like entity will contribute in this divergence.  Much like Proca's massive electromagnetism, one gets a Lorentz-like condition on the potential as a consequence.


\section{Universal Coupling of Ogievetsky-Polubarinov Mass Terms}

With the derivation of any arbitrary mass term from universal coupling now complete, it isn't strictly necessary to consider any specific cases.  Some definite examples might be helpful, however, to reduce the level of abstraction. First, consider the Ogievetsky-Polubarinov theories \cite{OP,OPmassive2}.  To fit the formalism, such theories must be adapted from Cartesian coordinates with imaginary time to formal general covariance with a flat metric \emph{tensor} $\eta_{\mu\nu}$ or some densitized and perhaps inverted relative thereof.  Thus polished, the theories are, apart from the familiar Einstein terms, 
\begin{equation}
\mathcal{L}_{ms} = \frac{m^2}{32 \pi G n} \left[ (4l-2) \sqrt{-g} - \frac{1}{n} (\tilde{g}^{\mu\nu} \tilde{\eta}_{\nu\alpha})^n \delta^{\alpha}_{\mu} \sqrt{-\eta} 
+ (2-4l + 4/n)\sqrt{-\eta} \right].
\end{equation} 
I use $l$ for western density weight, so $l=-p.$ 
Taking the derivative one gets
$$\lambda \frac{\delta S_{ms} }{\delta \tilde{\eta}^{\mu\nu} } |\tilde{g} = \frac{m^2}{\lambda n } \left[ \sqrt{-\eta} \tilde{\eta}_{\mu\alpha} (\tilde{g}^{\alpha\beta} \tilde{\eta}_{\beta\nu})^n  +  \frac{ \sqrt{-\eta} \tilde{\eta}_{\mu\nu} }{4 l -2 } \left( -\frac{1}{n}  (\tilde{g}^{\rho\sigma} \tilde{\eta}_{\sigma\alpha})^n \delta^{\alpha}_{\rho}  + 2 -4l + \frac{4}{n} \right) \right]. $$
This expression vanishes to zeroth order (as desired). It can be expanded to linear order in weight $l$ fields:  $\tilde{g}^{\mu\nu}= \tilde{\eta}^{\mu\nu} + \lambda \tilde{\gamma}^{\mu\nu} + \ldots.$    The resulting expansion agrees (to linear order, and up to a sign as expected) with 
$ \frac{\delta S_{fm}}{\delta \tilde{\gamma}^{\mu\nu}}  = -m^2 \sqrt{-\eta}  \tilde{\gamma}^{\alpha\beta} ( \tilde{\eta}_{\alpha \mu} \tilde{\eta}_{\beta \nu}  + b \tilde{\eta}_{\alpha \beta} \tilde{\eta}_{\mu \nu})$ when one sets $$b= -\frac{1}{n(4l-2)}.$$

Note that it was never necessary to write down  explicitly
 $A_{\mu\nu}^{\alpha\beta}(\eta, \gamma),$    $ \frac{\partial \tilde{g}^{\rho\sigma} }{ \partial \tilde{\gamma}^{\mu\nu} }$, or 
$ \frac{\partial \tilde{g}^{\rho\sigma} }{ \partial \tilde{\eta}^{\mu\nu}}.$ One could  attempt to do so after the fact. 
In this case simply altering the trace to isolate the gravitational potential does much of the job:
$$ \lambda n \tilde{\gamma}^{\phi\chi} = - \tilde{\eta}^{\phi\chi} + \tilde{\eta}^{\nu\chi} (\tilde{g}^{\phi\beta} \tilde{\eta}_{\beta\nu})^n,$$ 
which of course matches \cite[equation 113]{OP}, with arbitrary real coordinates in place of Cartesian coordinates and imaginary time.  The monomial in a power of the effective curved metric (inverted and weighted) $ \tilde{g}^{\phi\beta}$ makes this specific expression easy to solve:
$$   \tilde{g}^{\mu\nu} =   (\delta^{\mu}_{\alpha} + \lambda n \tilde{\gamma}^{\mu\beta} \tilde{\eta}_{\beta\alpha})^{\frac{1}{n}}  \tilde{\eta}^{\nu\alpha},$$ 
for which one can make a binomial series expansion  if desired
 \cite[equation 119b]{OP}.

But inverting the relationship between the potential and the metrics to express $\tilde{g}^{\mu\nu}$ in terms of $\tilde{\eta}^{\mu\nu}$ and $\tilde{\gamma}^{\mu\nu} $ is something that typically can \emph{only} be done perturbatively.  One probably wouldn't bother to try even in some cases where exact solution is possible. Even if one tried and succeeded, one would be working hard to achieve something that isn't clearly all that interesting (unless one believes that some relations of the gravitational potential to the metrics are fundamentally more correct---not merely more convenient---than others).  Thus a great deal of uninteresting labor is avoided and the full collection of every mass term such that \emph{there exists} some definition of the gravitational potential leading to it is found---and that collection includes pretty much every mass term.

%


\section{Pure Spin-$2$ Sum of dRGT-Hassan-Rosen Mass Terms from Ogievetsky-Polubarinov Family}

In the derivation above I have taken advantage of the option of using the same density weight $l$ for the field variables as appears in the theory at which I wanted to arrive.  Previously the medium (the density weight of the field) and the message (the resulting universally coupled field) have been closely linked \cite{MassiveGravity1,MassiveGravity2}.  That link no longer is required, which is good because not every massive gravity theory of interest has a preferred density weight or even a preference between covariant and contravariant indices.  Here is an example built around two old mass terms \cite{OP,ZuminoDeser}. In recent years three pure spin-$2$ massive gravities have been found, avoiding the Boulware-Deser-Tyutin-Fradkin ghost.  These theories admit linear combinations, thus giving a two-parameter family given fixed graviton mass.  Two of the three component theories fall within the Ogievetsky-Polubarinov family, but the techniques used by those authors do not permit adding two theories to give another in their family.  They were, of course, aware that even more general massive gravities were possible than they achieved using an expression linear in an arbitrary nonzero real powers of an arbitrarily densitized contravariant tensor \cite[p. 193]{OP}. It seems of interest, therefore, to study an arbitrarily weighted average of the two theories that were entertained by Ogievetsky and Polubarinov and that are now known to have pure spin-$2$.  As noted above, Maheshwari showed the pure spin-$2$ character of one of them long ago \cite{MaheshwariIdentity}, even before the Tyutin-Fradkin-Boulware-Deser announcement of a nonlinear ghost, but his paper did not receive any attention.  The two theories in question are the $n=\frac{1}{2}$, $p=-2$ theory built around the square root of the weight $2$ contravariant metric density (Maheshwari's pure spin-$2$) and the $ n= -\frac{1}{2}$, $p=0$ theory built around the square root $r_{\mu\nu}$ of the (weight $0$) covariant metric. 
Above I used $l$ for density weight in the usual western sense, so $l=-p.$
These theories are built around fields with different index placement (contravariant and covariant) and different density weight, so one would likely not consider them if one held to a One True Field Definition.  
   One  has
\begin{eqnarray} 
\mathcal{L}_{ms} = \frac{x}{x+y} \frac{m^2}{16 \pi G } \left[ 6 \sqrt{-g} - 2 (\sqrt{-g}^2 g^{\mu\nu} \eta_{\nu\alpha}/\sqrt{-\eta}^2)^{\frac{1}{2}} \delta^{\alpha}_{\mu} \sqrt{-\eta} 
+ 2\sqrt{-\eta} \right] \nonumber \\ 
-  \frac{y}{x+y} \frac{m^2}{16 \pi G } \left[ -2 \sqrt{-g} + 2 (\sqrt{-g}^0 g_{\mu\nu} \eta^{\nu\alpha}\sqrt{-\eta}^0)^{\frac{1}{2}} \delta_{\alpha}^{\mu} \sqrt{-\eta} 
-6 \sqrt{-\eta} \right] =  \nonumber \\
\frac{m^2}{(x+y)16 \pi G } \left[ (6x+2y) \sqrt{-g} - 2x ( g^{\mu\nu} \eta_{\nu\alpha})^{\frac{1}{2}} \delta^{\alpha}_{\mu} \sqrt{-g} 
- 2y (g_{\mu\nu} \eta^{\nu\alpha})^{\frac{1}{2}} \delta_{\alpha}^{\mu} \sqrt{-\eta} \right. \nonumber \\
\left. +(2x+6y) \sqrt{-\eta} \right].
\end{eqnarray} 
In calling this a weighted average, I do not mean to imply that $x$ and $y$ must be non-negative.  
This expression adds quantities naturally expressed using different weights and different valencies (covariant \emph{vs.} contravariant), so it provides a good occasion to notice that the choice of field variables (chosen as contravariant in this paper, but of almost any weight) is no longer tied to the resulting mass term. Making another more or less arbitrary choice for definiteness in this section, I use non-densitized ($l=0$) fields.

Finding the contribution of the full mass term to the on-shell remnants of the stress-energy tensor gives 
\begin{eqnarray}  \lambda \frac{\delta S_{ms} }{\delta {\eta}^{\mu\nu} } |g
=  \frac{2m^2}{\lambda(x+y)} \left[ x \sqrt{-g} \eta_{\mu\alpha}  ( g^{\alpha\beta} \eta_{\beta\nu})^{\frac{1}{2}}  \right. \nonumber \\ \left.
- y (g_{\mu\alpha} \eta^{\alpha\beta})^{\frac{1}{2}}   \eta_{\beta\nu} \sqrt{-\eta}  
+ y (g_{\alpha\beta} \eta^{\beta\rho})^{\frac{1}{2}} \delta^{\alpha}_{\rho} \sqrt{-\eta}  \eta_{\mu\nu}
-(x+3y) \sqrt{-\eta}  \eta_{\mu\nu}  \right].
\end{eqnarray} 
Expanding this expression to first order in contravariant weight-$0$ fields gives
\begin{eqnarray}
m^2 \sqrt{-\eta} (\gamma_{\mu\nu} - \gamma \eta_{\mu\nu}) + O(\lambda \gamma)^2,
\end{eqnarray}
which shows the appropriate Pauli-Fierz ghost-free mass term to linear order.  Equating this expression (up to a sign) with $\frac{\delta S_{fm}}{\delta \gamma^{\mu\nu}}$ works and yields $b=-1.$

If one undertakes  to invert the relationship to express $g^{\mu\nu}$ in terms of $\eta^{\mu\nu}$ and the gravitational potential $\gamma^{\mu\nu}$, then  one obtains a cubic matrix equation (cubic, linear, and constant terms) in the square root of $g_{\mu\nu}.$  It might possibly yield to exact treatment, unlike more general problems.  Working perturbatively instead, one already expects $g^{\mu\nu} = \eta^{\mu\nu} + \lambda \gamma^{\mu\nu} + \ldots,$ but what does one fill in for the ``\ldots''?  
There seems to be little to do other than allow terms of the forms $\gamma^{\mu\alpha} \gamma_{\alpha}^{\nu},$ $\gamma^{\mu\nu} \gamma,$ and $\gamma^2 \eta^{\mu\nu}$ with arbitrary coefficients.  Clearly this process will get less appealing at every order. This is the generic situation, whether or not this particular case admits an exact solution as a cubic equation.   It is much better to avoid such considerations and arrive directly at the conclusion that \emph{there exists} some definition of $\gamma^{\mu\nu}$ such that the resulting theory is universally coupled.  

 %


\section{Third (Non-trace) dRGT-Hassan-Rosen Pure Spin-$2$ Mass Term}

The previous  section used the sum of two theories of the formal cosmological constant + trace variety.  What if one considers theories involving a quadratic expression in some bimetric variables---whether chosen carefully to be ghost-free (such as the third, novel dRGT-Hassan-Rosen theory) or chosen on more formal grounds of simplicity?   Because universal coupling always works, it works for these sorts of theories also.  For  a mass term built out of the novel (non-Ogievetsky-Polubarinov-Zumino) pure spin-$2$ mass term,  $$\mathcal{L}_{ms} = \frac{m^2}{32\pi G} [6 \sqrt{-g} + \sqrt{-g}g^{\mu\nu} \eta_{\mu\nu} - \sqrt{-g} ([g^{\mu\nu}\eta_{\nu\alpha}]^{\frac{1}{2}} \delta_{\mu}^{\alpha})^2 + 6 \sqrt{-\eta}]$$
one (of course) satisfies the universal coupling condition for some as-yet unknown choice of the gravitational potential.
For consistency with the derivation above (rather than redoing it with covariant fields) one can choose contravariant fields, and let them have  weight $0$ for simplicity, and explore the process of inverting the expression $\gamma^{\mu\nu}=f(g^{\mu\nu}, \eta^{\mu\nu})$.  One gets
\begin{eqnarray}  -\lambda \frac{\delta S_{ms} }{\delta \eta^{\mu\nu} } |g =  -\lambda \frac{m^2}{32 \pi G} [ - \sqrt{-g} g^{\alpha\beta} \eta_{\alpha\mu} \eta_{\nu\beta}   - 3\sqrt{-\eta} \eta_{\mu\nu}     \nonumber  \\    + \sqrt{-g} (\sqrt{g^{\sigma\rho} \eta_{\rho\epsilon} } \delta^{\epsilon}_{\sigma} ) \sqrt{g^{\phi\chi} \eta_{\chi\mu} } \eta_{\nu\phi} ].  \end{eqnarray}
 The equation of this expression to $$ \frac{\delta S_{fm} }{\delta \gamma^{\mu\nu} } = -m^2 \sqrt{-\eta} (\gamma_{\mu\nu} - \gamma \eta_{\mu\nu} )$$ allows one to infer the relation of the $\gamma^{\mu\nu}$ to the flat and curved metrics.  That relation starts as $g^{\mu\nu}=\eta^{\mu\nu} + \lambda \gamma^{\mu\nu}+\ldots,$ but the terms ``\ldots'' have yet to be ascertained.  
Considered as an exact equation, it is a quadratic matrix equation (quadratic and constant terms) in the square root of $\sqrt{-g} g^{\mu\nu},$ but the quadratic term involves a matrix times its trace less the square of the  matrix, and so is not trivial to solve.  Perturbative treatment might be required.  
It should be straightforward to do this exercise using an arbitrary sum of the three ghost-free pure spin-$2$ mass terms, with even less hope of exact solution.  Thus one would never infer that these theories are universally coupled if one used only the traditional ``forward'' derivations instead of this paper's ``backward'' approach.


\section{Babak-Grishchuk Theories and Visser's Theory}

A few theories motivated by naive mathematical simplicity, taking the mass term as the square of a perturbation of a familiar metric-related quantity, have been  considered.  These theories also violate the $\sqrt{-g}+$ trace pattern. 
Thus these theories seemed not to be universally coupled even after $4$ one-parameter families of universally coupled theories were known.  Worse, these theories also have ghosts.
But one might wonder if naive (quadratic) mathematical simplicity has any benefits, benefits that might generalize in some interesting way?  
 The Babak-Grishchuk theories are built using $\mathfrak{g}^{\mu\nu} = \sqrt{-g}g^{\mu\nu}$ using an expression quadratic in the (de-densitized) deviation of this quantity from $\sqrt{-\eta} \eta^{\mu\nu}$ \cite[equation 15]{GrishchukMass}.  To avoid the proliferation of symbols needed only briefly, one can simply write the mass term in bimetric variables:
\begin{eqnarray}
\mathfrak{L}_{ms} = - \sqrt{-\eta} \frac{k_1}{2}\left[ \mathfrak{g}^{\rho\sigma} \eta_{\sigma\beta} \mathfrak{g}^{\beta\alpha} \eta_{\alpha\rho} /\sqrt{-\eta}^2  - 2 \mathfrak{g}^{\rho\sigma} \eta_{\rho\sigma}/\sqrt{-\eta} + 4 \right]  \nonumber \\ 
- \sqrt{-\eta} \frac{k_2}{2}\left[ (\mathfrak{g}^{\rho\sigma} \eta_{\rho\sigma}/\sqrt{-\eta})^2 - 8 \mathfrak{g}^{\rho\sigma} \eta_{\rho\sigma}/\sqrt{-\eta} + 16 \right].
\end{eqnarray} 
Although these theories have ghosts, it is surprisingly difficult to catch them doing anything spooky, at least classically \cite{GrishchukMass}.  But reality is not classical, so the expected quantum catastrophe makes them unappealing.  In any case, the task at hand is merely to illustrate how these theories fit within the realm of universal coupling. 
It is convenient to use weight $l=1$ fields, so $\tilde{g}^{\mu\nu} = \mathfrak{g}^{\mu\nu}.$ 
One finds that \begin{eqnarray}  \lambda \frac{ \delta S_{ms}  }{\delta \tilde{\eta}^{\mu\nu} }|\tilde{g}=    \lambda \frac{ \eta_{\mu\nu} }{2 \sqrt{-\eta} } \mathcal{L}_{ms} - \frac{ \lambda \sqrt{-\eta}}{2} [ - 2 k_1 \tilde{\eta}_{\mu\sigma} \tilde{g}^{\sigma\rho} \tilde{\eta}_{\rho\alpha} \tilde{g}^{\alpha\beta} \tilde{\eta}_{\beta\nu} \nonumber \\ - 2k_2 \tilde{g}^{\rho\sigma} \tilde{\eta}_{\rho\sigma} \tilde{\eta}_{\mu\alpha} \tilde{g}^{\alpha\beta} \tilde{\eta}_{\beta\nu} +(2k_1+8k_2)\tilde{\eta}_{\mu\rho} \tilde{g}^{\rho\sigma} \tilde{\eta}_{\sigma\nu} ]. \end{eqnarray}
Expanding to linear order using $\tilde{g}^{\mu\nu}=\tilde{\eta}^{\mu\nu} + \lambda \tilde{\gamma}^{\mu\nu} + \ldots,$ 
one sees that the term  $ \lambda \frac{ \eta_{\mu\nu} }{2 \sqrt{-\eta} } \mathcal{L}_{ms} $ does not contribute to lowest order and that the second term becomes  
$\sqrt{-\eta} 32\pi G(k_1 \tilde{\gamma}_{\mu\nu} + k_2 \gamma \tilde{\eta}_{\mu\nu}).$
Equating this expression to $-\frac{\delta S_{fm} }{\delta \tilde{\gamma}^{\mu\nu} }$ gives $k_1 = \frac{m^2}{32\pi G}$ and $k_2= \frac{b m^2}{32\pi G}.$ 

 If  one attempts to invert the relation to ascertain how $\tilde{g}^{\mu\nu}$ is built from $\tilde{\eta}^{\mu\nu}$ and that gravitational potential that yields universal coupling, one sees a quadratic matrix equation (quadratic, linear and constant terms) with two pieces in the quadratic term, somewhat worse than the third dRGT theory.  Once again it is best to avoid such inversion.

One could give a similar treatment to Visser's theory \cite{Visser}.  Altering the sign and normalization of the mass term by a factor of $-\frac{1}{32 \pi G}$ in equations 5 and 6, one finds that this theory also fits the universal coupling pattern.  Using weight $0$ fields $g^{\mu\nu}=\eta^{\mu\nu} + \lambda \gamma^{\mu\nu} +\ldots,$ one finds $b=-\frac{1}{2}.$


\section{How the Genericity of Universal Coupling Affects Theory Choice}

Universal coupling by itself is consistent with ghosts, tachyons, and whatever other bad behavior a massive graviton theory might display, so it is not a panacea.  But universal coupling arguably makes a theory better than any cousins that are not universally coupled but also have the same ghost presence or absence, and that fit the data equally well  \cite{FMS,DeserMass}.  (Such matters that now take one into the realm of effective field theory). Can one use universal coupling as a tie-breaker?  
 If the more crucial matters of handling the traditional worries about a ghost and/or vDVZ discontinuity have been addressed, one might  prefer any remaining candidates that are universally coupled---assuming that some theories are not!  
It appeared that only one theory, the Freund-Maheshwari-Schonberg theory, was universal coupled \cite{FMS,DeserMass} until recently--- putting universally coupled theories purely into the ghost category (too bad for universal coupling).  More recent work generalized universal coupling to more ghost theories \cite{MassiveGravity1} using the metric (or its inverse) with nearly any density weight, giving a ghost of tunable mass lighter than the spin-$2$
\cite{MassiveGravity1}.  Still more recent work using a tetrad (or its inverse) with nearly any density weight permitted theories
heavier ghosts and theories linearizing in the Fierz-Pauli ghost-free way \cite{MassiveGravity2}; these are part of the Ogievetsky-Polubarinov family \cite{OP,ZuminoDeser}.  As was shown recently \cite{deRhamGabadadze,HassanRosenNonlinear}
 (and in part long ago \cite{MaheshwariIdentity} without attracting attention),  two of those theories are nonlinearly ghost-free, as is a third new theory, or any mixture of the $3$.  Should one conclude that only two theories out of this 2-parameter family (at fixed spin-$2$ graviton mass) of ghost-free theories are universally coupled?  That would make universal coupling quite predictive, and correspondingly fragile.  But as shown above, universal coupling, far from having a unique result, in fact applies to any mass term, good or bad---ghost-free, ghost-filled, tachyonic, \emph{etc.}, given the flexibility in defining stress-energy and making field redefinitions employed above.  If universal makes a theory a little bit special, it turns out that all theories are a little bit special.


\section{Appendix:  Spinors}

It is expected that future work will explicitly include spinor fields.  Spinors have occasionally appeared in discussions of universal coupling derivations of Einstein's equations \cite{DeserSupergravity,Shirafuji,SliBimGRG}, though less often than such fundamental entities should.   Often the discussion reproduces the Weyl-Cartan claim \cite{WeylGravitationElectron,CartanSpinor} that coupling of spinors to the curved space-times implied by General Relativity requires the introduction of an orthonormal tetrad.  This claim was proven false by construction in the mid-1960s by Ogievetsky-Polubarinov the invention of nonlinear group realizations \cite{OP,OPspinor,GatesGrisaruRocekSiegel,BilyalovSpinors,PittsSpinor}.  Very roughly, one can take a tetrad, impose the very common symmetric gauge condition \cite{WoodardSymmetricTetrad}, and---this is the key step that is usually missed---then notice that one has a tetrad-free spinor formalism that makes perfectly good sense in its own right and even (at least in its bosonic sector) lacks the topological restrictions implied by a tetrad.  
Doubtless sometimes is \emph{convenient}  to represent the gravitational potential using an orthonormal tetrad, especially (but not only) if spinors are present, in order to avoid technically demanding nonlinearities \cite[p. 234]{GatesGrisaruRocekSiegel}, as the predecessor paper did \cite{MassiveGravity2}.  Including spinors along the lines of nonlinear (because metric-dependent) group realizations will require graviton-dependent field redefinitions of the spinor built around the square root of the flat metric  $\eta_{\mu\nu}$.  That is because the  square root of $\eta_{\mu\nu}$---roughly a symmetrized flat tetrad locally in admissible coordinates---would be rendered symmetric by one boost-rotation (fixing of the local Lorentz freedom), whereas the curved tetrad would be rendered symmetric by a different boost-rotation in general.  Thus a spinor built around the flat metric square root needs to be boosted and rotated in order to be built around the curved metric square root.  There will also be non-tensorial spinor-related parts of $\frac{\delta S }{\delta \eta_{\mu\nu} }$ and  $\frac{\delta S }{ \delta g _{\mu\nu} }$, so one cannot assume tensor properties as usual from the notation, a point that was not made clear in (\cite{SliBimGRG}).  

There has appeared a fairly detailed technical and historical discussion of the issue of spinors in coordinates to subsume them as far as possible within classical nonlinear geometric objects (components relative to a coordinate system and a transformation rule \cite{PittsSpinor,Nijenhuis,Tashiro1,Tashiro2,SzybiakLie,SzybiakCovariant}.  Most strikingly, it turns out that the coordinate atlas has subtleties if the nonlinear realizations use an indefinite signature matrix such as $diag(-1,1,1,1),$ as one would want to do for space-time. This is an interpretation of one of the jobs of Bilyalov's pivoting matrix $T$.  
(The issues involving the square root of the metric in graviton mass terms are analogous \cite{DeffayetSymmetricTetrad}.) To that story I can now add that a symmetric gauge condition (the first step toward spinors in coordinates \emph{via} nonlinear group realizations) was imposed at lowest order already by Bronstein in 1936 \cite{BronsteinWave} and that a fair amount of the apparatus of spinors as nonlinear group realizations appeared in Bryce (Seligman) DeWitt's dissertation \cite[p. 66]{DeWittDissertation}.
  Thus the already known footnote 7 of (\cite{DeWittSpinor}) reflects a more substantial earlier treatment.  DeWitt seems not to have heartily embraced the conceptual innovation of nonlinear group realizations as Ogievetsky and Polubarinov later would \cite{OP,OPspinor}.  Nonetheless one can now suggest his dissertation as perhaps the first serious step toward nonlinear group realizations.
%

%

\section{Acknowledgments}
I thank Stanley Deser,  Andrew Tolley and Fawad Hassan for discussing Maheshwari's work, Alex Blum for acquainting me with the Bronstein paper and the Seligman (DeWitt) dissertation, and Jeremy Butterfield for discussion.  This work was funded by the John Templeton Foundation, grant \#38761.
%


\end{document}